\documentclass[aps,preprint]{revtex4}%
\usepackage{amsfonts}
\usepackage{amsmath}
\usepackage{amssymb}
\usepackage{graphicx}%
\begin{document}
\title{{\LARGE Pryce-Hoyle Tensor in a Combined Einstein-Cartan-Brans-Dicke Model}}
\author{Marcelo Samuel Berman$^{1}$}
\affiliation{$^{1}$Instituto Albert Einstein/Latinamerica\ - Av. Candido Hartmann, 575 -
\ \# 17}
\affiliation{80730-440 - Curitiba - PR - Brazil - email: msberman@institutoalberteinstein.org}
\date{(Original: 10 January, 2008; Last Version: 14 July, 2008)}

\begin{abstract}
In addition to introducing matter injection through a scalar field determined
by Pryce-Hoyle tensor, we also combine it with a BCDE
(Brans-Dicke-Einstein-Cartan) theory \ with lambda-term developed earlier by
Berman(2008), for inflationary scenario. It involves a variable cosmological
constant, which decreases with time, jointly with energy density, cosmic
pressure, shear, vorticity, and Hubble's parameter, while the scale factor,
total spin and scalar field increase exponentially. The post-inflationary
fluid resembles a perfect one, though total spin grows, but not the angular
speed (Berman, 2007d). The Pryce-Hoyle tensor, which can measured by the
number of injected particles per unit proper volume and time, as well as shear
and vorticity, can be neglected in the aftermath of inflation ("no-hair").



\end{abstract}
\maketitle

\begin{center}

{\LARGE Pryce-Hoyle Tensor in a Combined Einstein-Cartan-Brans-Dicke Model }

\bigskip Marcelo Samuel Berman

\bigskip
\end{center}

{\large I. INTRODUCTION}

\bigskip

\bigskip The purpose of this paper, is to show that, when exponential
inflation is turned on, in the Universe, eventual shear, vorticity, or matter
injection (originated from a Pryce-Hoyle term), which may have originated in
the very early Universe, are completely erased ("no-hair") by the enormous
expansion which represents this phase. We had arrived to this conclusion, when
the Pryce-Hoyle tensor is absent, in two previous papers (Berman, 2007c;
2008). We shall find the same results, for the present case.

\bigskip

If the Universe is rotating, i.e., if it has a non-zero total-spin, the
left-handed creation characteristic (Barrow and Silk, 1983) of the Universe
would be explained; this would also attach a meaning to parity violation, and
thus, according to the teachings of Feynman et al (1965), it would explain the
matter-antimatter asymmetry, plus the Pioneer anomaly (Berman, 2007d), and
neutrinos left handed spin. Berman(2008c,d), has shown that Robertson-Walker's
metric includes a "hidden" state of rotation plus expansion.

\bigskip

\bigskip Since the advent of Modern Cosmology (Weinberg, 1972; 2008), two
different kinds of models turned-out of the cosmologists' brains: the
big-bang, and the stationary. Now that we believe about inflation, as a
possible phase of the early Universe, it is difficult to detach the
exponential inflationary big-bang model, from the exponential stationary one.
(for inflation, see Kolb and Turner, 1990; Weinberg, 2008). For instance, the
reader may check the books by Narlikar for a description of both kinds of
models (Narlikar, 1983; 1993). While matter injection has been put forward
with the stationary model, it seems feasible that one could think of such
subject, by introducing, into the energy-momentum-tensor employed in the
big-bang picture, a Pryce-Hoyle component. This matter injection tensor leads
to the so-called C-field, and when we apply to Robertson-Walker's metric, the
expanding Universe obeys field equations tantamount to adding the term
\ $\frac{1}{2}\kappa f$ $\dot{\lambda}^{2}$\ \ \ to both the energy density
and cosmic pressure equations.

\bigskip

Dirac proposed that a time-varying gravitational "constant", could be needed
in order to explain some "coincidences" of a cosmological nature (Dirac,
1938). Later, Brans and Dicke \ (1961) included this variation, as a scalar
field, in order to accommodate Machian ideas (Berman, 2007; 2007a; 2008a).
Scalar fields in 5-D gravitational theories reduce to 4-D theories with a
cosmological constant, but the idea of a universal scalar field is present in
modern gravitational "scalar-tensor" theories (Berman, 2007a). String theories
also introduce "dilaton fields", which are also present in gravitation
counterparts. Berman (2008b) has even calculated energy and momenta of
rotating dilaton black holes, which were depicted as possible astrophysical objects.

\bigskip

We shall first review Pryce-Hoyle theory (Section II), then we shall deal
(Section III) with the combined torsion plus scalar-tensor gravitational
theory, as presented in our recent paper (Berman, 2008), and afterwards, we
derive the cosmological solution for a lambda-Universe in an inflationary
scenario with matter injection, where the fluid is endowed with shear and
vorticity\ (Section IV). In the final Section V, we comment the solution just derived.

\bigskip

{\Large II. REVIEW OF PRYCE-HOYLE THEORY}

\bigskip

\bigskip When steady-state theory was devised (Narlikar, 1993), the stationary
exponential Universe led to creation of matter: consider a proper three-volume,

\bigskip

$V\propto e^{3Ht}$ \ \ \ \ \ \ \ \ \ \ \ \ \ \ \ \ .

\bigskip

Then, we obtain,

\bigskip

$\frac{\dot{V}}{V}=3H$ \ \ \ \ \ \ \ \ \ \ \ \ \ \ \ \ .

\bigskip

Consider the constant energy density \ \ $\rho=\rho_{0}$\ \ \ . The amount of
matter within the volume \ \ $V$\ \ \ would grow like,

\bigskip

$\dot{M}=3HV\rho$ \ \ \ \ \ \ \ \ \ \ \ \ \ \ \ \ \ \ ,

\bigskip

so that, the rate of creation of matter per unit volume would be something like,

\bigskip

$Q=3H\rho$ \ \ \ \ \ \ \ \ \ \ \ \ \ \ \ \ \ .

\bigskip

\bigskip As Berman(2008a) has reported, Hoyle (1948), inspired by the above
calculation, introduced, in Cosmology, this additional term towards the energy
momentum tensor, originated by a scalar field, responsible for matter
injection. This field, due to Pryce, Hoyle and Narlikar (Narlikar, 1993; Hoyle
and Narlikar, 1963; Berman and Marinho Jr., 1996), is represented by:

\bigskip

$T^{\mu\nu}=T_{M}^{\mu\nu}-f$ $\left(  \lambda^{\mu}\lambda^{\nu}-\frac{1}%
{2}g^{\mu\nu}\lambda^{\alpha}\lambda_{\alpha}\right)  $
\ \ \ \ \ \ \ \ \ \ \ \ \ \ \ \ \ , \ \ \ \ \ \ \ \ \ \ \ \ \ \ \ \ \ \ \ \ \ \ \ \ \ \ \ \ \ \ \ \ \ \ \ \ \ \ \ \ \ \ \ (1)

\bigskip

where, \ \ $T_{M}^{\mu\nu}$\ \ stands for the normal matter energy-momentum
tensor, and \ $f$\ \ is a constant, while \ $\lambda_{\mu}$\ \ is a vector
given by:

\bigskip

$\lambda_{\mu}=\frac{\partial\lambda}{\partial x^{\mu}}=\left(  0,0,0,\dot
{\lambda}\right)  $\ \ \ \ \ \ \ \ \ \ \ \ \ \ \ \ . \ \ \ \ \ \ \ \ \ \ \ \ \ \ \ \ \ \ \ \ \ \ \ \ \ \ \ \ \ \ \ \ \ \ \ \ \ \ \ \ \ \ \ \ \ \ \ \ \ \ \ \ \ \ \ \ \ \ \ \ \ \ \ \ (2)

\bigskip

Einstein's equations are kept like:

\bigskip

$G^{\mu\nu}=-\kappa T^{\mu\nu}$ \ \ \ \ \ \ \ \ \ \ \ \ \ \ \ \ \ \ \ \ \ . \ \ \ \ \ \ \ \ \ \ \ \ \ \ \ \ \ \ \ \ \ \ \ \ \ \ \ \ \ \ \ \ \ \ \ \ \ \ \ \ \ \ \ \ \ \ \ \ \ \ \ \ \ \ \ \ \ \ \ \ \ \ \ \ \ \ \ \ \ \ \ (3)

\bigskip

There is an additional relation,

\bigskip

$n=j_{;\mu}^{\mu}$ \ \ \ \ \ \ \ \ \ \ \ \ , \ \ \ \ \ \ \ \ \ \ \ \ \ \ \ \ \ \ \ \ \ \ \ \ \ \ \ \ \ \ \ \ \ \ \ \ \ \ \ \ \ \ \ \ \ \ \ \ \ \ \ \ \ \ \ \ \ \ \ \ \ \ \ \ \ \ \ \ \ \ \ \ \ \ \ \ \ \ \ \ \ \ \ \ \ \ \ \ \ (4)

\bigskip

which stands for the number of particles injected per unit of proper 4-volume,
the particle current being represented by \ \ $j^{\mu}$ \ \ . \ For
Robertson-Walker's metric,

\bigskip

$ds^{2}=dt^{2}-\frac{R^{2}(t)}{\left[  1+\left(  \frac{kr^{2}}{4}\right)
\right]  ^{2}}d\sigma^{2}$ \ \ \ \ \ \ \ \ \ \ \ \ \ \ , \ \ \ \ \ \ \ \ \ \ \ \ \ \ \ \ \ \ \ \ \ \ \ \ \ \ \ \ \ \ \ \ \ \ \ \ \ \ \ \ \ \ \ \ \ \ \ \ \ \ \ \ \ \ \ \ \ \ \ \ (5)

\bigskip

where,

\bigskip

$d\sigma^{2}=dx^{2}+dy^{2}+dz^{2}$ \ \ \ \ \ \ \ \ \ \ \ \ , \ \ \ \ \ \ \ \ \ \ \ \ \ \ \ \ \ \ \ \ \ \ \ \ \ \ \ \ \ \ \ \ \ \ \ \ \ \ \ \ \ \ \ \ \ \ \ \ \ \ \ \ \ \ \ \ \ \ \ \ \ \ \ \ \ \ \ \ \ \ (6)

\bigskip

we find the following field equations:

\bigskip

$\kappa\rho=3\left(  \frac{\dot{R}}{R}\right)  ^{2}+3\frac{k}{R^{2}}+\frac
{1}{2}\kappa f\dot{\lambda}^{2}$ \ \ \ \ \ \ \ \ \ \ \ ,
\ \ \ \ \ \ \ \ \ \ \ \ \ \ \ \ \ \ \ \ \ \ \ \ \ \ \ \ \ \ \ \ \ \ \ \ \ \ \ \ \ \ \ \ \ \ \ \ \ \ \ \ \ \ \ \ \ \ \ \ (7)\bigskip

\bigskip

and,

\bigskip

$\kappa p=-2\frac{\ddot{R}}{R}-\left(  \frac{\dot{R}}{R}\right)  ^{2}-\frac
{k}{R^{2}}+\frac{1}{2}\kappa f\dot{\lambda}^{2}$ \ \ \ \ \ \ \ \ \ \ \ .
\ \ \ \ \ \ \ \ \ \ \ \ \ \ \ \ \ \ \ \ \ \ \ \ \ \ \ \ \ \ \ \ \ \ \ \ \ \ \ \ \ \ \ \ \ \ \ \ \ \ \ (8)\bigskip

\bigskip

Additionally we have an equation for matter injection proper,

\bigskip

$\ddot{\lambda}+3\dot{\lambda}\frac{\dot{R}}{R}=f^{-1}n=f^{-1}j_{;\mu}^{\mu}$
\ \ \ \ \ \ \ \ \ \ \ \ \ \ \ \ \ \ \ \ \ \ \ \ \ ,
\ \ \ \ \ \ \ \ \ \ \ \ \ \ \ \ \ \ \ \ \ \ \ \ \ \ \ \ \ \ \ \ \ \ \ \ \ \ \ \ \ \ \ \ \ \ \ \ \ \ \ (9)\bigskip

\bigskip

where \ $n(t)$\ \ stands for the number of particles injected per unit of
proper volume and proper time.

\bigskip

\bigskip It was Narlikar, in 1973, that noticed that the C-field, could be
either considered as representing continuous matter injection or "explosive"
big-bang \ paraphernalia.

\bigskip

For instance, let us work a simple case.

From the field equations, with,

\bigskip

$R=R_{0}e^{Ht}$ \ \ \ \ \ \ \ \ \ \ \ \ \ . \ \ \ \ \ \ \ \ ( $R_{0}%
=$\ \ constant )\ \ \ \ \ \ \ \ \ \ \ \ \ \ \ \ \ \ \ \ \ \ \ \ \ \ \ \ \ \ \ \ \ \ \ \ \ \ \ \ \ \ \ \ (10)\ 

\bigskip

$\ddot{\lambda}=0$ \ \ \ \ \ \ \ \ \ .

\bigskip

$\dot{\lambda}=\frac{n}{3fH}=$ \ constant\ \ \ \ \ \ \ \ \ \ \ \ \ . \ \ \ \ \ \ \ \ \ \ \ \ \ \ \ \ \ \ \ \ \ \ \ \ \ \ \ \ \ \ \ \ \ \ \ \ \ \ \ \ \ \ \ \ \ \ \ \ \ \ \ \ \ \ \ \ \ \ \ \ \ \ \ \ (11)

\bigskip

$\kappa\rho=3H^{2}+\kappa\frac{f}{2}\dot{\lambda}^{2}=3H^{2}+\frac{\kappa
}{18f}H^{-2}n^{2}=$ \ constant\ \ \ \ \ \ \ \ \ . \ \ \ \ \ \ \ \ \ \ \ \ \ \ \ \ \ \ \ \ \ \ \ (12)

\bigskip

$\kappa p=-3H^{2}+\frac{\kappa}{18f}H^{-2}n^{2}=$
\ constant\ \ \ \ \ \ \ \ \ \ \ \ \ \ \ . \ \ \ \ \ \ \ \ \ \ \ \ \ \ \ \ \ \ \ \ \ \ \ \ \ \ \ \ \ \ \ \ \ \ \ \ \ \ (13)

\bigskip

It is supposed, in the above model, that \ $n$\ \ is
constant!!!\ Whitrow-Randall's relation would also applicable in this case, so
that we can call such model as Machian.

\bigskip

{\Large \bigskip III. REVIEW OF THE COMBINED BCDE THEORY}

\bigskip

Berman(2007b), examined the time behavior of shear and vorticity in a
lambda-Universe, for inflationary models, in a Brans-Dicke framework. The
resulting scenario is that exponential inflation smooths the fluid, in order
to become a nearly perfect one after the inflationary period. In a subsequent
paper (Berman, 2008), a similar inflationary scenario was examined with the
inclusion of torsion, \textit{\`{a} la} Einstein-Cartan, when a scalar field
of Brans-Dicke origin, is included, \ along with a Cosmological lambda-term.
Again, with suitable constants, the model performed adequately, and, while
total spin grew, along with scale-factor and scalar field, all other
characteristics decreased, and the post-inflationary fluid, resembled a
perfect one.

\bigskip{\large \bigskip}\bigskip Einstein-Cartan's gravitational theory,
though not bringing vacuum solutions different than those in General
Relativity theory, has an important r\^{o}le, by tying macrophysics, through
gravitational and electromagnetic phenomena (i.e., involving constants
\ $G$\ \ and $c$\ ), with microphysics, though Planck's constant, involving
spin originated by torsion. Intrinsic angular momentum was introduced by
Cartan as a Classical quantity (Cartan, 1923) before it was introduced as a
Quantum Theory element, around 1925. Of course, spin is important in the
Quantum Theory of particles. However, spin has taken part of Classical Field
Theory for a long time, and Cosmological models were treated as early as 1973
(Trautman, 1973). Einstein-Cartan Theory is the simplest Poincar\'{e} gauge
theory of gravity, in the frame of which, the gravitational field is described
by means of curvature and torsion, the sources being energy-momentum and spin
tensors. It is important to stress that torsion can be originated by spin but
not necessarily vice-versa. We mean that Einstein-Cartan's theory, is not the
only possible framework for a theory involving spin. Just look at General
Relativity Theory, which may include angular momentum phenomena, even without
evoking torsion.

\bigskip

\bigskip Though it was in the past, supposed that, due to spin,
Robertson-Walker's metric might not be representative of Physical reality in a
torsioned spacetime, recent papers recalled the approach shown by us in
several papers (Berman, 1990; 1991), on how anisotropic Bianchi-I models in
Einstein-Cartan's theory could be reduced to Robertson-Walker's prototype, by
defining overall, deceleration parameters, and scale-factors; we did the same
thing, with \ other papers dealing with anisotropic models in GRT and BD
theories [ for GRT see (Berman, 1988; Berman and Som, 1989 b); for BD theory
see (Berman and Som,1989) ].\ \ On the other hand, Berman and Som (2007) have
shown that, slight deviations from Robertson-Walker's metric, changing it to a
Bianchi-I metric, are enough to produce the anisotropic phenomena, like
entropy production, or other ones; this is a clue to the possibility of
considering overall scale-factors and deceleration parameters, etc, in the
Raychaudhuri's equation for Einstein-Cartan's Cosmology, without worrying with
any anisotropy, which becomes implicit in the equations of \ Raychaudhuri's
book (Raychaudhuri, 1979). The essential modification of General Relativistic
Bianchi-I cosmology, when we carry towards Einstein-Cartan's, resides, when
field equations are explicited, in that the normal energy momentum tensor
components \ \ $T_{1}^{1}$\ \ , \ $T_{2}^{2}$\ \ \ \ and\ \ $T_{3}^{3}$\ \ are
subtracted by a term \ \ $S^{2}$\ \ , while \ \ $T_{0}^{0}$\ \ is added by
\ \ $S^{2}$\ \ . Of course, there appear also non-diagonal \ $S-$\ dependent
terms: for instance, \ $T_{3}^{2}$\ \ and \ \ $T_{2}^{3}$\ \ depend linearly
with \ \ $S^{32}$\ . \ \ In our treatment of the Einstein-Cartan-Brans-Dicke
theory, the field equations are obviously satisfied, but we have short-cutted
the derivations, like we have done in the previous paper (Berman, 2007b),
which also conforms with the field equations of that case (Brans-Dicke theory
with lambda).\ \ The off-diagonal energy momentum components are null, for a
Robertson-Walker's framework.

\bigskip

\bigskip It is generally accepted that scalar tensor cosmologies play a
central r\^{o}le in the present view of the very early Universe (Berman,
2007). The cosmological "constant", which represents quintessence, may be a
time varying entity, whose origin remounts to Quantum theory(Berman, 2007a),
but see also a possible Classical explanation for lambda in Berman (2008e).
The first, and most important scalar tensor theory was devised by Brans and
Dicke(1961), which is given in the "Jordan's frame". Afterwards, Dicke(1962)
presented a new version of the theory, in the "Einstein's frame", where the
field equations resembled Einstein's equations, but time, length, and inverse
mass, were scaled by a factor \ $\phi^{-\frac{1}{2}}$\ \ where \ $\phi
$\ \ stands for the scalar field. Then, the energy momentum tensor
\ \ $T_{ij}$\ \ is augmented\ by a new term $\Lambda_{ij}$\ , so that:

\bigskip

$G_{ij}=-8\pi G\left(  T_{ij}+\Lambda_{ij}\right)  $\ \ \ \ \ \ \ \ \ \ , \ \ \ \ \ \ \ \ \ \ \ \ \ \ \ \ \ \ \ \ \ \ \ \ \ \ \ \ \ \ \ \ \ \ \ \ \ \ \ \ \ \ \ \ \ \ \ \ \ \ \ (14)

\bigskip

where \ \ $G_{ij}$\ \ stands for Einstein's tensor. The new energy tensor
quantity, is given by:

\bigskip

$\Lambda_{ij}=\frac{2\omega+3}{16\pi G\phi^{2}}\left[  \phi_{i}\phi_{j}%
-\frac{1}{2}G_{ij}\phi_{k}\phi^{k}\right]  $ \ \ \ \ \ \ \ \ \ \ \ \ . \ \ \ \ \ \ \ \ \ \ \ \ \ \ \ \ \ \ \ \ \ \ \ \ \ \ \ \ \ \ \ \ \ \ \ \ \ (15)

\bigskip

In the above, \ $\omega$\ \ is the coupling constant. The other equation is:

\bigskip

$\square\log\phi=\frac{8\pi G}{2\omega+3}T$\ \ \ \ \ \ , \ \ \ \ \ \ \ \ \ \ \ \ \ \ \ \ \ \ \ \ \ \ \ \ \ \ \ \ \ \ \ \ \ \ \ \ \ \ \ \ \ \ \ \ \ \ \ \ \ \ \ \ \ \ \ \ \ \ \ \ \ \ \ \ \ (16)

\bigskip

where \ $\square$\ \ is the generalized d'Alembertian, and $T=T_{i}^{i}%
$\ \ \ .\ \ It is useful to remember that the energy tensor masses are also
scaled by \ $\phi^{-\frac{1}{2}}$\ \ .

\bigskip

For\ the Robertson-Walker's flat metric,

\bigskip

$ds^{2}=dt^{2}-\frac{R^{2}(t)}{\left[  1+\left(  \frac{kr^{2}}{4}\right)
\right]  ^{2}}d\sigma^{2}$ \ \ \ \ \ \ \ \ \ \ \ \ \ \ , \ \ \ \ \ \ \ \ \ \ \ \ \ \ \ \ \ \ \ \ \ \ \ \ \ \ \ \ \ \ \ \ \ \ \ \ \ \ \ \ \ \ \ (17)

\bigskip

where \ \ $k=0$\ \ and \ $d\sigma^{2}=dx^{2}+dy^{2}+dz^{2}$\ \ .

\bigskip

The field equations now read, in the alternative Brans-Dicke
reformulation(Raychaudhuri, 1979):

\bigskip

$\frac{8\pi G}{3}\left(  \rho+\frac{\Lambda}{\kappa}+\rho_{\lambda}\right)
=H^{2}\equiv\left(  \frac{\dot{R}}{R}\right)  ^{2}$ \ \ \ \ \ \ \ \ \ \ \ . \ \ \ \ \ \ \ \ \ \ \ \ \ \ \ \ \ \ \ \ \ \ \ \ \ \ \ \ \ \ \ \ \ \ \ \ \ (18)

\bigskip

$-8\pi G\left(  p-\frac{\Lambda}{\kappa}+\rho_{\lambda}\right)  =H^{2}%
+\frac{2\ddot{R}}{R}$ \ \ \ \ \ \ \ \ \ \ \ \ \ \ \ \ . \ \ \ \ \ \ \ \ \ \ \ \ \ \ \ \ \ \ \ \ \ \ \ \ \ \ \ \ \ \ \ \ \ (19)

\bigskip

In the above, we have: \ 

$\bigskip$

$\rho_{\lambda}=\frac{2\omega+3}{32\pi G}\left(  \frac{\dot{\phi}}{\phi
}\right)  ^{2}=\rho_{\lambda0}\left(  \frac{\dot{\phi}}{\phi}\right)  ^{2}$
\ \ \ \ \ \ \ . \ \ \ \ \ \ \ \ \ \ \ \ \ \ \ \ \ \ \ \ \ \ \ \ \ \ \ \ \ \ \ \ \ \ \ \ \ \ \ \ \ (20)

\bigskip

From the above equations (18), (19) and (20) we obtain:

\bigskip

$\frac{\ddot{R}}{R}=-\frac{4\pi G}{3}\left(  \rho+3p+4\rho_{\lambda}%
-\frac{\Lambda}{4\pi G}\right)  $ \ \ \ \ \ \ \ \ \ \ \ \ \ . \ \ \ \ \ \ \ \ \ \ \ \ \ \ \ \ \ \ \ \ \ \ \ \ \ \ \ \ \ \ \ (21)

\bigskip

Relation (21) represents Raychaudhuri's equation for a perfect fluid. By the
usual procedure, we would find the Raychaudhuri's equation in the general
case, involving shear ($\sigma_{ij}$) and vorticity ($\varpi_{ij}$); the
acceleration of the fluid is null for the present case, and then we find:

\bigskip

$3\dot{H}+3H^{2}=2\left(  \varpi^{2}-\sigma^{2}\right)  -4\pi G\left(
\rho+3p+4\rho_{\lambda}\right)  +\Lambda$ \ \ \ \ \ \ \ \ , \ \ \ \ \ \ \ \ \ \ \ (22)\ \ \ 

\bigskip

where \ $\Lambda$\ \ stands for a cosmological "constant". As we are mimicking
Einstein's field equations, $\Lambda$\ \ in (22) stands like it were a
constant (see however, Berman, 2007, 2007a, 2006b, 2006c). Notice that, \ when
we impose that the fluid is not accelerating, this means that the
quadri-velocity is tangent to the geodesics, i.e., the only interaction is gravitational.

\bigskip

\bigskip When Raychaudhuri's equation is calculated for non-accelerated fluid,
taken care of Einstein-Cartan's theory, combined with Brans-Dicke theory, the
following equation was found by us, based on the original calculation for
Einstein-Cartan's theory by Raychaudhuri (1979):

\bigskip

$3\dot{H}+3H^{2}=2\varpi^{2}-2\sigma^{2}-4\pi G\left(  \rho+3p+4\rho_{\lambda
}\right)  +\Lambda+128\pi^{2}S^{2}$ \ \ \ \ \ \ , \ \ (23)\ \ \ 

\bigskip

where \ \ $S$\ \ stands for the spin density contents of the fluid, where we
have omitted a term like

\bigskip

$\varpi S=\varpi_{ik}S^{ik}+\varpi^{ik}S_{ik}$%
\ \ \ \ \ \ \ \ \ \ \ \ \ \ \ \ \ \ \ \ \ , \ \ \ \ \ \ \ \ \ \ \ \ \ \ \ \ \ \ \ \ \ \ \ \ \ \ \ \ \ \ \ \ \ \ \ \ \ \ \ \ \ \ \ \ \ \ \ \ \ (24)

\bigskip

which is to be included in the pressure and energy density terms, by a re-scaling.

\bigskip

The introduction of Pryce-Hoyle tensor, as far as pressure and energy-density
are concerned, \ can be done by the addition of the term \ $\frac{1}{2}\kappa
f$ $\dot{\lambda}^{2}$\ \ , as exposed above. The Raychaudhuri's \ equation
would have the following form for a non-accelerating fluid:

\bigskip

$3\dot{H}+3H^{2}=2\varpi^{2}-2\sigma^{2}-\frac{\kappa}{2}\left(  \rho
+3p+4\rho_{\lambda}+\kappa f\text{ }\dot{\lambda}^{2}\right)  +128\pi^{2}%
S^{2}+\Lambda$ \ \ . \ \ \ \ \ \ (25)

\bigskip

For inflation, we shall impose, that:

\bigskip

$3H^{2}=\Lambda$ \ \ \ \ \ \ \ \ \ \ \ \ \ \ \ . \ \ \ \ \ \ \ \ \ \ \ \ \ \ \ \ \ \ \ \ \ \ \ \ \ \ \ \ \ \ \ \ \ \ \ \ \ \ \ \ \ \ \ \ \ \ \ \ \ \ \ \ \ \ \ \ \ \ \ \ \ \ \ \ \ \ \ \ \ \ \ \ (26)

\bigskip

It is important to stress, that relation (23) is the same general relativistic
equation, with the additional spin term, which transforms it into
Einstein-Cartan's equation. When we work a combined Einstein-Cartan's and
Brans-Dicke theory (BCDE theory), we would need to calculate the new field
equations for the combined theory.

\bigskip By employing the total action (Sabbata and Gasperini, 1985),

$\bigskip$

$L=\int d^{4}x\sqrt{-g}\left[  \pounds _{m}\left(  \psi,\triangledown
\psi,g\right)  -\frac{1}{2\chi}R\left(  g,\partial g,Q\right)  \right]  $
\ \ \ \ \ \ \ \ \ \ \ , \ \ \ \ \ \ \ \ \ \ \ \ \ (27)

\bigskip

where the matter Lagrangian contains torsion because the connection is not
symmetric, and $\chi$\ \ is the coupling constant, both for curvature and
torsion, and when we perform independent variations with respect to \ $\psi
$\ \ , \ $g_{\mu\nu}$\ \ and \ $Q_{\mu\nu}^{\alpha}$\ \ \ ; the last the
one\ is the torsion tensor,

\bigskip

$Q_{\alpha\beta}^{\mu}=\frac{1}{2}\left(  \Gamma_{\alpha\beta}^{\mu}%
-\Gamma_{\beta\alpha}^{\mu}\right)  $ \ \ \ \ \ \ \ . \ \ \ \ \ \ \ \ \ \ \ \ \ \ \ \ \ \ \ \ \ \ \ \ \ \ \ \ \ \ \ \ \ \ \ \ \ \ \ \ \ \ \ \ \ \ \ \ \ \ \ \ \ \ \ \ (28)

\bigskip

We find, the Einstein tensor,

\bigskip

$G^{\mu\nu}-\hat{\triangledown}_{\alpha}\left(  T^{\mu\nu\alpha}-T^{\nu
\alpha\mu}+T^{\alpha\mu\nu}\right)  =\chi T^{\mu\nu}$ \ \ \ \ \ \ , \ \ \ \ \ \ \ \ \ \ \ \ \ \ \ \ \ \ \ \ \ \ \ \ \ \ \ \ \ \ \ \ (29)

\bigskip

where,

\bigskip

$T^{\mu\nu\alpha}=\chi S^{\mu\nu\alpha}$ \ \ \ \ \ \ \ \ \ \ \ .\ \ \ \ \ \ \ \ \ \ \ \ \ \ \ \ \ \ \ \ \ \ \ \ \ \ \ \ \ \ \ \ \ \ \ \ \ \ \ \ \ \ \ \ \ \ \ \ \ \ \ \ \ \ \ \ \ \ \ \ \ \ \ \ \ \ (30)

\bigskip

We have defined,

\bigskip

$\hat{\triangledown}_{\alpha}\equiv\triangledown_{\alpha}+2Q_{\alpha
}=\triangledown_{\alpha}+2Q_{\alpha\nu}^{\nu}$ \ \ \ \ \ \ \ \ \ \ , \ \ \ \ \ \ \ \ \ \ \ \ \ \ \ \ \ \ \ \ \ \ \ \ \ \ \ \ \ \ \ \ \ \ \ \ \ \ \ \ \ (31)

\bigskip

while the modified torsion tensor,

\bigskip

$T_{\mu\nu}^{\alpha}=Q_{\mu\nu}^{\alpha}+\delta_{\mu}^{\alpha}Q_{\nu}%
-\delta_{\nu}^{\alpha}Q_{\mu}$ \ \ \ \ \ \ \ \ \ \ \ \ \ \ . \ \ \ \ \ \ \ \ \ \ \ \ \ \ \ \ \ \ \ \ \ \ \ \ \ \ \ \ \ \ \ \ \ \ \ \ \ \ \ \ \ (32)

\bigskip

{\Large \bigskip IV. SOLUTION FOR BCDE THEORY WITH MATTER INJECTION}

\bigskip

\bigskip From the prior Sections, we now are able to write the resulting
equations for a perfect fluid, which can be inferred from Raychaudhuri (1979):

\bigskip$-8\pi\left[  \frac{1}{2}\kappa f\dot{\lambda}^{2}+p\right]  =\left[
\text{Brans-Dicke alternative Riemann tensor \ }G_{i}^{i}\text{\ }\right]
+256\pi^{2}S^{2}$ \ \ , \ \ \ (33)

\bigskip$8\pi\left[  \frac{1}{2}\kappa f\dot{\lambda}^{2}+\rho\right]
=\left[  \text{Brans-Dicke alternative Riemann tensor \ }G_{0}^{0}%
\text{\ }\right]  +256\pi^{2}S^{2}$ \ \ \ \ \ , \ \ (34)

\bigskip

It is important to acknowledge, that the above field equations should be
applied into the pseudo-General Relativistic equations, i.e., the Brans-Dicke
alternative (unconventional) framework. A plausibility reasoning that
substitutes an otherwise lengthy \ calculation, is the following: the term
with spin, as well as it is added to the other general relativistic terms in
equation (23), should be added equally to equation \ (22), because this is the
Brans-Dicke equation in a general relativistic format. This equation is
written in the unconventional format (Dicke, 1962), i.e., the alternative
system of equations. We could not write so simply equation\ (23)\ if the terms
in it were those of conventional Brans-Dicke theory.

\bigskip

It is important to stress, that \ $\lambda\left(  t\right)  $\ \ still has to
obey the conservation equation (22)

\bigskip

Consider now exponential inflation, like we find in Einstein's theory
(Weinberg, 2008):

\bigskip

$R=R_{0}e^{Ht}$\ \ \ \ \ \ \ \ , \ \ \ \ \ \ \ \ \ \ \ \ \ \ \ \ \ \ \ \ \ \ \ \ \ \ \ \ \ \ \ \ \ \ \ \ \ \ \ \ \ \ \ \ \ \ \ \ \ \ \ \ \ \ \ \ \ \ \ \ \ \ \ \ \ \ \ \ \ \ \ \ \ (35)

\bigskip

and, \ as usual in General Relativity inflationary models,

\bigskip

$\Lambda=3H^{2}$ \ \ \ \ \ \ \ \ \ \ . \ \ \ \ \ \ \ \ \ \ \ \ \ \ \ \ \ \ \ \ \ \ \ \ \ \ \ \ \ \ \ \ \ \ \ \ \ \ \ \ \ \ \ \ \ \ \ \ \ \ \ \ \ \ \ \ \ \ \ \ \ \ \ \ \ \ \ \ \ \ \ \ (36)

\bigskip

For the time being, $H$ \ is just a constant, defined by \ $H=\frac{\dot{R}%
}{R}$\ \ . We shall see, when we go back to conventional\ \ Brans-Dicke
theory, that \ $H$\ \ is not the Hubble's constant.

\bigskip

From (35), we find $H=H_{0}=$\ constant.\ 

\bigskip

A\ solution of Raychaudhuri's equation\ (23), would be the following:

\bigskip

\bigskip$\sigma=\sigma_{0}e^{-\frac{\beta}{2}t}$ \ \ \ ;

\bigskip

$\varpi=\varpi_{0}e^{-\frac{\beta}{2}t}$ \ \ \ ;

\bigskip

$\rho=\rho_{0}e^{-\beta t}$ \ \ \ ;

\bigskip

$p=p_{0}e^{-\beta t}$ \ \ \ ; \ \ \ \ \ \ \ \ \ \ \ \ \ \ \ \ \ \ \ \ \ \ \ \ \ \ \ \ \ \ \ \ \ \ \ \ \ \ \ \ \ \ \ \ \ \ \ \ \ \ \ \ \ \ \ \ \ \ \ \ \ \ \ \ \ \ \ \ \ \ \ \ \ \ \ (37)

\bigskip

$\phi=\phi_{0}e^{-\frac{\beta}{2}\sqrt{A}\text{ \ }e^{-\frac{\beta}{2}t}}$ \ \ \ \ \ \ \ \ \ .

\bigskip

$\Lambda=\Lambda_{0}=$ constant.

\bigskip

\bigskip$\lambda=-\frac{\lambda_{0}}{2H}e^{-2Ht}$ \ \ \ \ \ \ \ \ \ \ \ \ \ \ .

\bigskip$n=f$ $H\lambda_{0}e^{-2Ht}$ \ \ \ \ \ \ \ \ \ \ \ \ .

\bigskip$S_{U}=SR^{3}=s_{0}R_{0}^{3}e^{Ht}$ \ \ \ \ \ \ \ \ \ \ \ \ .

\bigskip In the above, \ $\lambda_{0}$ $\ $,$\ \ \sigma_{0}$\ , \ $\phi_{0}%
$\ ,\ \ $p_{0}$\ \ , \ \ $\rho_{0}$\ , $\beta$\ , \ $s_{0}$\ and \ $R_{0}$\ ,
are constants, and, \ \ $S_{U}$\ \ stands for the total spin of the Universe,
whose spin density equals,

\bigskip

$S=s_{0}e^{-\frac{\beta}{2}t}=s_{0}e^{-2Ht}$\ \ \ \ \ \ \ \ \ , \ \ \ \ \ \ \ \ \ \ \ \ \ \ \ \ \ \ \ \ \ \ \ \ \ \ \ \ \ \ \ \ \ \ \ \ \ \ \ \ \ \ \ \ \ \ \ \ \ \ \ \ (38)

\bigskip

while,

\bigskip

$\beta=4H$\ \ \ \ \ . \ \ \ \ \ \ \ \ \ \ \ \ \ \ \ \ \ \ \ \ \ \ \ \ \ \ \ \ \ \ \ \ \ \ \ \ \ \ \ \ \ \ \ \ \ \ \ \ \ \ \ \ \ \ \ \ \ \ \ \ \ \ \ \ \ \ \ \ \ \ \ \ \ \ \ \ \ (39)

\bigskip

\bigskip The ultimate justification for this solution is that one finds a good
solution in the conventional units theory, and that the Universe must expand.

\bigskip

When we return to conventional units, we retrieve the following corresponding solution:

\bigskip

$\bar{R}=R_{0}\phi^{\frac{1}{2}}e^{Ht}$ \ \ \ \ \ \ \ \ ; \ \ \ \ \ \ \ \ \ \ \ \ \ \ \ \ \ \ \ \ \ \ \ \ \ \ \ \ \ \ \ \ \ \ \ \ \ \ \ \ \ \ \ \ \ \ \ \ \ \ \ \ \ \ \ 

\bigskip

$\bar{\rho}=\rho_{0}\phi^{-2}e^{-\beta t}$\ \ \ \ \ \ \ \ \ ;

\bigskip

$\bar{p}=p_{0}\phi^{-2}e^{-\beta t}=\left[  \frac{p_{0}}{\rho_{0}}\right]
\bar{\rho}$\ \ \ \ \ \ \ \ \ ;

\bigskip\ \ \ \ \ \ \ \ \ \ \ \ \ \ \ \ \ \ \ \ \ \ \ \ \ \ \ \ \ \ \ \ \ \ \ \ \ \ \ \ \ \ \ \ \ \ \ \ \ \ \ \ \ \ \ \ \ \ \ \ \ \ \ \ \ \ \ \ \ \ \ \ \ \ \ \ \ \ \ \ \ \ \ \ \ \ \ \ \ \ \ \ \ \ \ \ \ \ \ (40)

$\bar{\sigma}=\sigma\phi^{-\frac{1}{2}}$ \ \ \ \ \ \ \ \ \ \ \ \ \ \ ;

\bigskip

$\bar{\varpi}=\varpi\phi^{-\frac{1}{2}}$ \ \ \ \ \ \ \ \ \ \ \ \ \ \ ;

\bigskip

$\bar{\Lambda}=\Lambda_{0}\phi^{-1}$\ \ \ \ \ \ \ \ \ \ \ \ \ ;

\bigskip

\bigskip$\bar{\phi}=\phi=\phi_{0}e^{-\frac{\beta}{2}\sqrt{A}\text{
\ }e^{-\frac{\beta}{2}t}}$ \ \ \ \ \ \ \ \ \ ;

\bigskip

$\overset{\cdot}{\bar{\lambda}}$ $=\phi^{-1}\dot{\lambda}$ \ \ \ \ \ \ \ \ \ \ \ \ \ \ \ \ \ ;

\bigskip

$\bar{n}=n\phi^{-2}$ \ \ \ \ \ \ \ \ \ \ \ \ \ \ .

\bigskip

\bigskip We also have,

\bigskip$\bar{S}_{U}=S_{U}=s_{0}R_{0}^{3}e^{Ht}$
\ \ \ \ \ \ ,\ \ \ \ \ in\ \ \ \ \ $c=1$\ \ \ units\ \ \ \ . \ \ \ \ \ \ \ \ \ \ \ \ \ \ \ \ \ \ \ \ (41)

\bigskip

As we promised to the reader, $H$\ is not the Hubble's constant. Instead, we find:

\bigskip

\bigskip$\bar{\Lambda}=\Lambda_{0}$ $\phi_{0}^{-1}$ $e^{\frac{\beta}{2}%
\sqrt{A}\text{ }e^{-\frac{\beta}{2}t}}$ \ \ \ \ ; \ \ \ \ \ \ \ \ \ \ \ \ \ \ \ \ \ \ \ \ \ \ \ \ \ \ \ \ \ \ \ \ \ \ \ \ \ \ \ \ \ \ \ \ \ \ \ \ \ \ \ \ \ \ \ \ (42)

\bigskip

$\bar{\rho}=\rho_{0}$ $\phi_{0}^{-2}$ $e^{\beta\left[  \sqrt{A}\text{
\ }e^{-\frac{\beta}{2}\text{ }t}-\text{ }t\right]  }$ \ \ \ \ \ \ ;\ \ \ \ \ \ \ \ \ \ \ \ \ \ \ \ \ \ \ \ \ \ \ \ \ \ \ \ \ \ \ \ \ \ \ \ \ \ \ \ \ \ \ \ \ \ \ \ \ \ (43)

\bigskip

$\bar{p}=p_{0\text{ }}\phi_{0}^{-2}$ $e^{\beta\left[  \sqrt{A}\text{
\ }e^{-\frac{\beta}{2}\text{ }t}-\text{ }t\right]  }$ \ \ \ \ \ \ ;\ \ \ \ \ \ \ \ \ \ \ \ \ \ \ \ \ \ \ \ \ \ \ \ \ \ \ \ \ \ \ \ \ \ \ \ \ \ \ \ \ \ \ \ \ \ \ \ \ \ \ (44)

\bigskip

$\bar{R}=R_{0}$ $\phi_{0}^{-\frac{1}{2}}$ $e^{\left[  H\text{ }t\text{ }%
-\frac{1}{4}\beta\text{ }\sqrt{A}\text{ \ }e^{-\frac{\beta}{2}\text{ }%
t}\right]  }$ \ \ \ \ \ \ ;\ \ \ \ \ \ \ \ \ \ \ \ \ \ \ \ \ \ \ \ \ \ \ \ \ \ \ \ \ \ \ \ \ \ \ \ \ \ \ \ \ \ \ \ (45)

\bigskip

\bigskip$\bar{\sigma}=\sigma_{0}$ $\phi_{0}^{-\frac{1}{2}}$ $e^{-\frac{1}%
{2}\beta\left[  \text{ }t\text{ }-\frac{1}{2}\text{ }\sqrt{A}\text{
\ }e^{-\frac{\beta}{2}\text{ }t}\right]  }$ \ \ \ \ \ \ ,\ \ \ \ \ \ \ \ \ \ \ \ \ \ \ \ \ \ \ \ \ \ \ \ \ \ \ \ \ \ \ \ \ \ \ \ \ \ \ \ \ \ \ \ (46)

\bigskip

$\bar{\varpi}=\varpi_{0}$ $\phi_{0}^{-\frac{1}{2}}$ $e^{-\frac{1}{2}%
\beta\left[  \text{ }t\text{ }-\frac{1}{2}\text{ }\sqrt{A}\text{ \ }%
e^{-\frac{\beta}{2}\text{ }t}\right]  }$ \ \ \ \ \ \ ,\ \ \ \ \ \ \ \ \ \ \ \ \ \ \ \ \ \ \ \ \ \ \ \ \ \ \ \ \ \ \ \ \ \ \ \ \ \ \ \ \ \ (47)

\bigskip

$\overset{\cdot}{\bar{\lambda}}$ $=\dot{\lambda}$ $\phi_{0}^{-1}e^{\frac
{\beta}{2}\left[  \sqrt{A}\text{ \ }e^{-\frac{\beta}{2}\text{ }t}\right]  }$
\ \ \ \ \ \ \ \ \ \ \ \ \ \ \ , \ \ \ \ \ \ \ \ \ \ \ \ \ \ \ \ \ \ \ \ \ \ \ \ \ \ \ \ \ \ \ \ \ \ \ \ \ \ \ \ \ \ \ \ \ \ (48)

\bigskip

and,

\bigskip

$\bigskip\bar{H}=H$ $\phi_{0}^{-\frac{1}{2}}$ $e^{\frac{1}{4}\beta\sqrt
{A}\text{ \ }e^{-\frac{\beta}{2}\text{ }t}}>0$ \ \ \ \ \ \ .\ \ \ \ \ \ \ \ \ \ \ \ \ \ \ \ \ \ \ \ \ \ \ \ \ \ \ \ \ \ \ \ \ \ \ \ \ \ \ \ \ \ \ \ \ \ \ \ \ \ (49)

\bigskip

From relation (48), we may calculate, by integration, the value for
\ $\lambda(t)$ \ :\ 

\bigskip

$\bar{\lambda}(t)=\phi_{0}^{-1}\lambda_{0}\int e^{2H\left[  \sqrt{A}%
e^{-2Ht}-\text{ }t\right]  }dt$ \ \ \ \ \ \ \ \ \ \ \ \ .

\bigskip

Altogether, we find,

\bigskip

$\bar{n}=f$ $\lambda_{0}\phi_{0}^{-2}H$ $e^{2H\left[  2\sqrt{A}e^{-2Ht}-\text{
}t\right]  }$ \ \ \ \ \ \ \ \ \ \ \ \ \ \ .

\bigskip

\bigskip The fluid obeys a perfect gas equation of state. It represents a
radiation phase, if we impose,

\bigskip

$p_{0}=\frac{1}{3}\rho_{0}$ \ \ \ \ \ \ \ \ \ \ \ \ \ . \ \ \ \ \ \ \ \ \ \ \ \ \ \ \ \ \ \ \ \ \ \ \ \ \ \ \ \ \ \ \ \ \ \ \ \ \ \ \ \ \ \ \ \ \ \ \ \ \ \ \ \ \ \ \ \ \ \ \ \ \ \ \ \ \ \ \ \ (50)

\bigskip

\bigskip Returning to Raychaudhuri's equation, we have the following condition
to be obeyed by the constants:

\bigskip

$\sigma_{0}^{2}-\varpi_{0}^{2}=-2\pi G\left[  \rho_{0}+3p_{0}+4\rho_{\lambda
0}+\kappa f\text{ }\lambda_{0}^{2}\right]  +64\pi^{2}s_{0}^{2}$
\ \ \ \ \ \ \ . \ \ \ \ \ \ \ \ \ \ \ \ \ \ \ \ \ (51)

\bigskip

{\Large \bigskip V. ANALYSIS AND COMMENTS OF THE RESULTS}

\bigskip We now investigate the limit when \ \ $t\longrightarrow\infty$\ \ of
the above formulae, having in mind that, by checking that limit, \ we will
know which ones increase or decrease with time; of course, we can not stand
with an inflationary period unless it takes only an extremely small period of
time. Remember that \ \ \ $\beta=4H>0$\ \ \ .

\bigskip

We find:

\bigskip

$\lim\limits_{t\longrightarrow\infty}\bar{H}=H\phi_{0}^{-1/2}$ \ \ \ \ ;\ \ \ 

\bigskip

$\lim\limits_{t\longrightarrow\infty}\bar{R}=\infty$ \ \ \ \ ;

\bigskip

$\lim\limits_{t\longrightarrow\infty}\bar{\sigma}=\lim
\limits_{t\longrightarrow\infty}\bar{\varpi}=0$ \ \ \ \ ;

\bigskip

$\lim\limits_{t\longrightarrow\infty}\bar{\rho}=\lim\limits_{t\longrightarrow
\infty}\bar{p}=0$ \ \ \ \ \ \ ;

\bigskip

$\lim\limits_{t\longrightarrow\infty}\bar{\Lambda}=\Lambda_{0}\phi_{0}^{-1}$ \ \ \ \ \ \ \ \ \ \ \ ;

\bigskip

$\lim\limits_{t\longrightarrow\infty}\bar{\phi}=\phi_{0}$ \ \ \ \ \ \ \ \ \ \ ;

\bigskip

\bigskip$\lim\limits_{t\longrightarrow\infty}\bar{S}_{U}=\infty$ \ \ \ \ \ \ ;

\bigskip

\bigskip$\lim\limits_{t\longrightarrow\infty}\bar{n}=0$ \ \ \ \ \ \ .

\bigskip

By comparing the above limits, \ with the limit \ $t\rightarrow0$\ \ , as we
can check, the scale factor, total spin, and the scalar field, are
time-increasing, while all other elements of the model, namely, vorticity,
shear, Hubble's parameter, energy density, cosmic pressure, the number of
particles injected per unit proper volume and proper time, and the
cosmological term, as described by the above relations, decay with time. This
being the case, shear and vorticity are decaying, so that, after inflation, we
retrieve\ \ a nearly perfect fluid: \ inflation has the peculiarity of
removing shear, and vorticity, but not spin, from the model. It has to be
remarked, that pressure and energy density obey a perfect gas equation of
state. The graceful exit from the inflationary period towards the early
Universe radiation phase, is attained with condition (50). We have found a
solution that is entirely compatible with the Brans-Dicke counterpart (Berman,
2007c). The total spin of the Universe grows, but the angular velocity does
not (Berman, 2007d). By the end of inflation, the number \ $\bar{n}$\ \ of
injected particles has practically died away, so that, for present day
Universe, the Pryce-Hoyle tensor\ has negligible effect: we have a kind of
cosmological result. It is a remarkable novel result, that the input of matter
injection, shear and vorticity, do not place any footprint into the final
state of the Universe, in the aftermath of inflation ("no-hair").

\bigskip\bigskip

{\Large ACKNOWLEDGEMENTS}

\bigskip

The author thanks his intellectual mentors, Fernando de Mello Gomide and M. M.
Som, and also to Marcelo Fermann Guimar\~{a}es, Nelson Suga, Mauro Tonasse,
Antonio F. da F. Teixeira, and for the encouragement by Albert, Paula and Geni.

\bigskip

\bigskip{\Large References}

\bigskip Barrow, J.D; Silk, J. (1983) - \textit{The Left Hand of Creation: The
Origin and Evolution of Expanding Universe, }Basic Books, New York.

Berman,M.S. (1988) - GRG \textbf{20}, 841.

Berman, M.S. (\bigskip1990) - Nuovo Cimento, \textbf{105B}, 1373.

Berman, M.S. (1991) - GRG, \textbf{23}, 1083.

\bigskip\bigskip Berman,M.S. (2006b) - \textit{Energy of Black-Holes and
Hawking's Universe \ }in \textit{Trends in Black-Hole Research, }Chapter
5\textit{.} Edited by Paul Kreitler, Nova Science, New York.

Berman,M.S. (2006c) - \textit{Energy, Brief History of Black-Holes, and
Hawking's Universe }in \textit{New Developments in Black-Hole Research},
Chapter 5\textit{.} Edited by Paul Kreitler, Nova Science, New York.

Berman,M.S. (2007) - \textit{Introduction to General Relativistic and Scalar
Tensor Cosmologies}, Nova Science, New York.

Berman,M.S. (2007a) - \textit{Introduction to General Relativity and the
Cosmological Constant Problem}, Nova Science, New York.

Berman,M.S. (2007c) - \textit{Shear and Vorticity in Inflationary Brans-Dicke
Cosmology with Lambda-Term, }Astrophysics and Space Science, \textbf{310, }205.

Berman,M.S. (2007d) - \textit{The Pioneer Anomaly and a Machian Universe,
}Astrophysics and Space Science, \textbf{312}, 275.

Berman,M.S. (2008) - \textit{Shear and Vorticity in a Combined
Einstein-Cartan-Brans-Dicke Inflationary Lambda-Universe, }Astrophysics and
Space Science, \textbf{314,} 79-82.

Berman,M.S. (2008a) - \textit{A Primer in Black-Holes, Mach's Principle and
Gravitational Energy}, Nova Science, New York.

Berman,M.S. (2008b) - \textit{Energy and Angular Momentum of a Dilaton Black
Holes, }Revista Mexicana de Astronom\'{\i}a y Astrof\'{\i}sica, \textbf{44}, -
. To appear in October issue.

Berman,M.S. (2008c) - \textit{A General Relativistic Rotating Evolutionary
Universe, }Astrophysics and Space Science, \textbf{314}, 319-321.

Berman,M.S. (2008d) - \textit{A General Relativistic Rotating Evolutionary
Universe - Part II, }Astrophysics and Space Science, to appear.

Berman,M.S. (2008e) - \textit{On the Rotational and Machian Properties of the
Universe, }\ Los Alamos Archives, http://arxiv.org/abs/physics/0610003 .

Berman,M.S.; Marinho Jr., R.M. (1996) - Nuovo Cimento, \textbf{B111}, 1279.

Berman,M.S.; Som, M.M. (1989) - Nuovo Cimento, \textbf{103B},N.2, 203.

Berman,M.S.;Som, M.M. (1989 b) - GRG , \textbf{21}, 967-970.

Berman,M.S.;Som, M.M. (2007) - \textit{Natural Entropy Production in an
Inflationary Model for a Polarized Vacuum}, Astrophysics and Space Science,
\textbf{310, }277. Los Alamos Archives:
http://www.arxiv.org/abs/physics/0701070 .

\bigskip Brans, C.; Dicke, R.H. (1961) - Physical Review, \textbf{124}, 925.

Cartan, E. (1923) - Ann. Ec. Norm. Sup., \textbf{40,} 325.

Dicke, R.H. (1962) - Physical Review, \textbf{125}, 2163.

Dirac, P.A.M. (1938) - Proceedings of the Royal Society \textbf{165A}, 199.

Feynman, R.P.; Leighton, R.B.; Sands, M. (1965) - \textit{The Feynman Lectures
on Physics.} Volume 1, Addison-Wesley, Reading.

Hoyle, F. (1948) - MNRAS \textbf{108}, 372.

Hoyle, F.; Narlikar, J. (1963) - Proc. Royal Society, \textbf{273A}, 1.

Kolb, E.W.; Turner, M.S. (1990) - \textit{The Early Universe, }Addison-Wesley, Reading.

Narlikar, J. (1983) - \textit{Introduction to Cosmology}, Bartlett, Boston.

Narlikar, J. (1993) - \textit{Introduction to Cosmology}, Second Edition, CUP, Cambridge.

\bigskip Raychaudhuri, A. K. (1979) - \textit{Theoretical Cosmology, }Oxford
University Press, Oxford.

Sabatta, V. de;\ Gasperini, M. (1985) - \textit{Introduction to Gravitation,
}World Scientific, Singapore.

Trautman, A. (1973) - Nature (Physical Science), \textbf{242,} 7.

Weinberg, S. (1972) - \textit{Gravitation and Cosmology, }Wiley, New York.

Weinberg, S. (2008) - \textit{Cosmology, }OUP, Oxford.

\end{document}